\documentclass[11pt,twoside]{article}
\usepackage{newpasp}

\newcommand{\HI}{\hbox{{\rm H}\kern 0.1em{\sc i}}}
\newcommand{\MgII}{\hbox{{\rm Mg}\kern 0.1em{\sc ii}}}
\newcommand{\FeII}{\hbox{{\rm Fe}\kern 0.1em{\sc ii}}}
\newcommand{\CIV}{\hbox{{\rm C}\kern 0.1em{\sc iv}}}
\newcommand{\NV}{\hbox{{\rm N}\kern 0.1em{\sc v}}}
\newcommand{\OVI}{\hbox{{\rm O}\kern 0.1em{\sc vi}}}
\newcommand{\kms}{\hbox{km~s$^{-1}$}}

\newcommand{\Lya}{\hbox{{\rm Ly}\kern 0.1em$\alpha$}}
\newcommand{\CIVdblt}{{\rm C}\kern 0.1em{\sc iv}~$\lambda\lambda 1548, 1550$}
\newcommand{\MgIIdblt}{{\rm Mg}\kern 0.1em{\sc ii}~$\lambda\lambda 2796, 2803$}

\index{Churchill, C. W.}

\begin{document}

\title{Multiphase Gas in Intermediate Redshift Galaxies}
\author{Chris Churchill$^{1}$, Rick Mellon$^{1}$, Jane Charlton$^{1}$,
\& Buell Januzzi$^{2}$}
\affil{$^{1}$The Pennsylvania State University, $^{2}$NOAO} 


\begin{abstract}
We present 40 quasar absorption line systems at intermediate redshifts
($z \sim  1$), with  focus on one  of the most  kinematically complex
known, as examples of how  the unique capabilities of space--based and
ground--based  facilities  can  be  combined  to  glean  much  broader
insights into astrophysical systems.
\end{abstract}




\section{Hubble and the More Complete Picture}

Within  the  field  of  quasar absorption  lines,  one  long--standing
question is how the halos and ISM of earlier epoch galaxies compare or
relate, in an evolutionary sense,  to those of the present epoch.  The
look--back time  to $z=1$ covers  well more than  half the age  of the
universe.   Furthermore,  spectral  and  morphological  properties  of
absorbing galaxies  are accessible with present  day ground--based and
spaced--based  observatories  (Steidel,  Dickinson, \&  Persson  1994;
Steidel  1998).    Thus,  absorption  line   studies  at  intermediate
redshifts provide  an opportunity to examine the  gaseous evolution of
galaxies.

The ISM and  halos of local galaxies are  comprised of many ionization
phases, including  diffused ionized gas, extended  coronae, and denser
low ionization regions  often located in front of  shock fronts (e.g.\
Dahlem 1998).  In  absorption, simultaneous study of both  the low and
high ionization phases  in our Galaxy have been  required to constrain
the ionization mechanisms, chemical abundance variations, and the dust
properties (e.g.\ Savage \& Sembach 1996).

A  significant obstacle  in the  face of  rapid progress  with studies
employing absorption lines, however, is that the strongest transitions
of the  cosmologically most abundant elements  lie in the  far to near
ultraviolet   (UV)   portion    of   the   electromagnetic   spectrum.
Fortunately,  at $z\sim1$,  the near  UV transitions,  which  are most
often associated with neutral and low ionization ions\footnote{Meaning
ions  with  ionization potentials  in  the range  of  a  few to  $\sim
30$~eV.}, are redshifted into the visible.  Thus, they can be observed
from the ground  with large aperture telescopes.  However,  the far UV
transitions,   associated   with    moderate   and   high   ionization
ions\footnote{Meaning those with ionization potentials ranging between
$\sim  30$  and $\sim  50$~eV  and  between  $\sim 50$  and  $140$~eV,
respectively.}, are  redshifted to  the near UV;  a study of  the high
ionization  component requires a  spaced--based telescope,  i.e.\ {\it
HST}.  The  {\it HST\/} archive is  rich with $R=1300$  FOS spectra of
quasars,  the majority  due to  the  QSO Absorption  Line Key  Project
(Bahcall et~al.\ 1993).

\begin{figure}[ht]
\plotfiddle{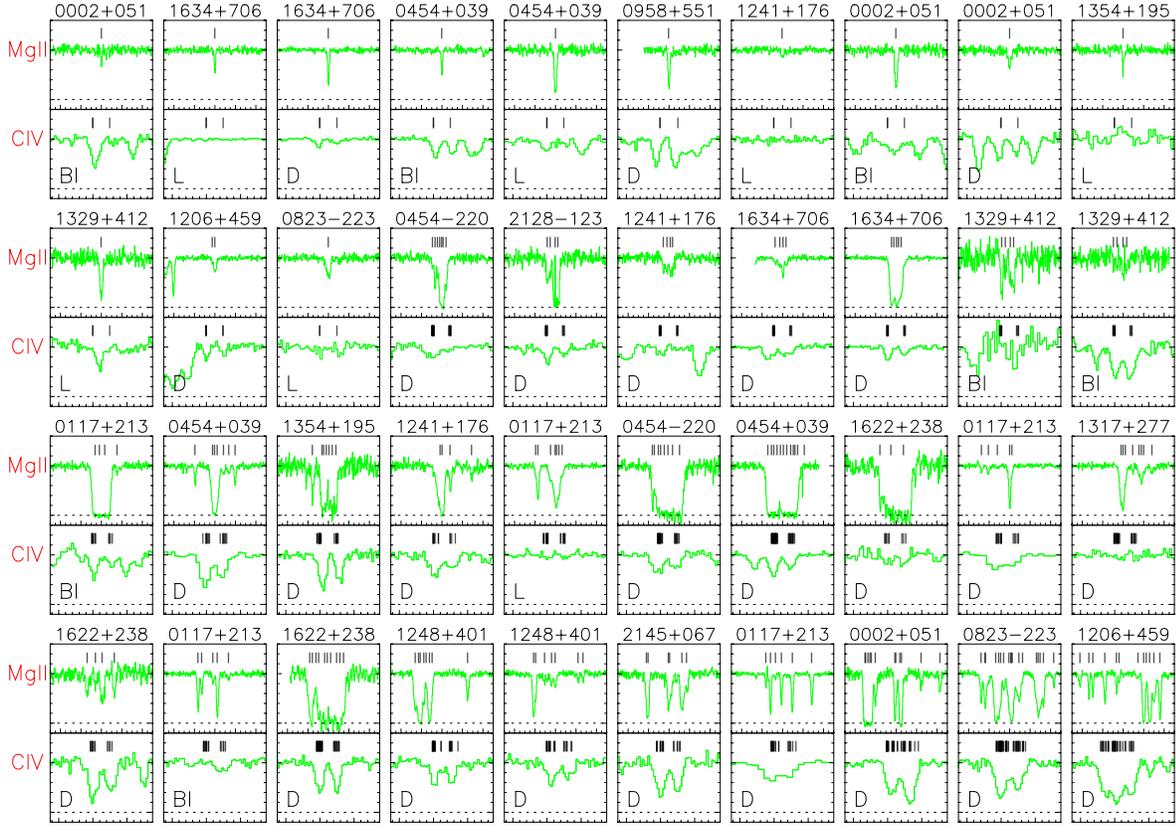}{4.in}{0.}{70.}{70.}{-284}{-78}
\caption
{ In order  of increasing {\MgII} kinematic spread  are the HIRES/Keck
sample of systems for which  FOS/{\it HST\/} spectra of the {\CIVdblt}
doublet were available from the archive.  See text for details.}
\end{figure}

\section{The C\kern 0.1em {\footnotesize\bf IV}--Mg\kern 0.1em 
{\footnotesize\bf II}  Kinematics Connection: Multiphase Gas}

We used  HIRES/Keck spectra  ($R\sim 6$~{\kms}) and  archival FOS/{\it
HST\/}  spectra  ($R\sim  230$~{\kms})  to place  constraints  on  the
ionization and multiphase distribution  of absorbing gas at $z=0.4$ to
$z=1$.   In  Figure~1,  we  present  {\MgII}~$\lambda  2796$  and  the
{\CIVdblt}  doublet for  each of  40 systems  (note that  the velocity
scale for {\MgII} is 500~{\kms} and for {\CIV} is 3000~{\kms}).  Ticks
above  the HIRES  spectra give  the  velocities of  the Voigt  profile
{\MgII} sub--components and ticks above the FOS data give the expected
location  of these  components  for the  {\CIV}  doublet.  The  labels
``D'',  ``L'',   and  ``Bl''  denote  detection,   limit,  and  blend,
respectively.   The  systems  are  presented in  order  of  increasing
{\MgII} kinematic spread from the upper left to lower right.

\begin{figure}[p]
\plotfiddle{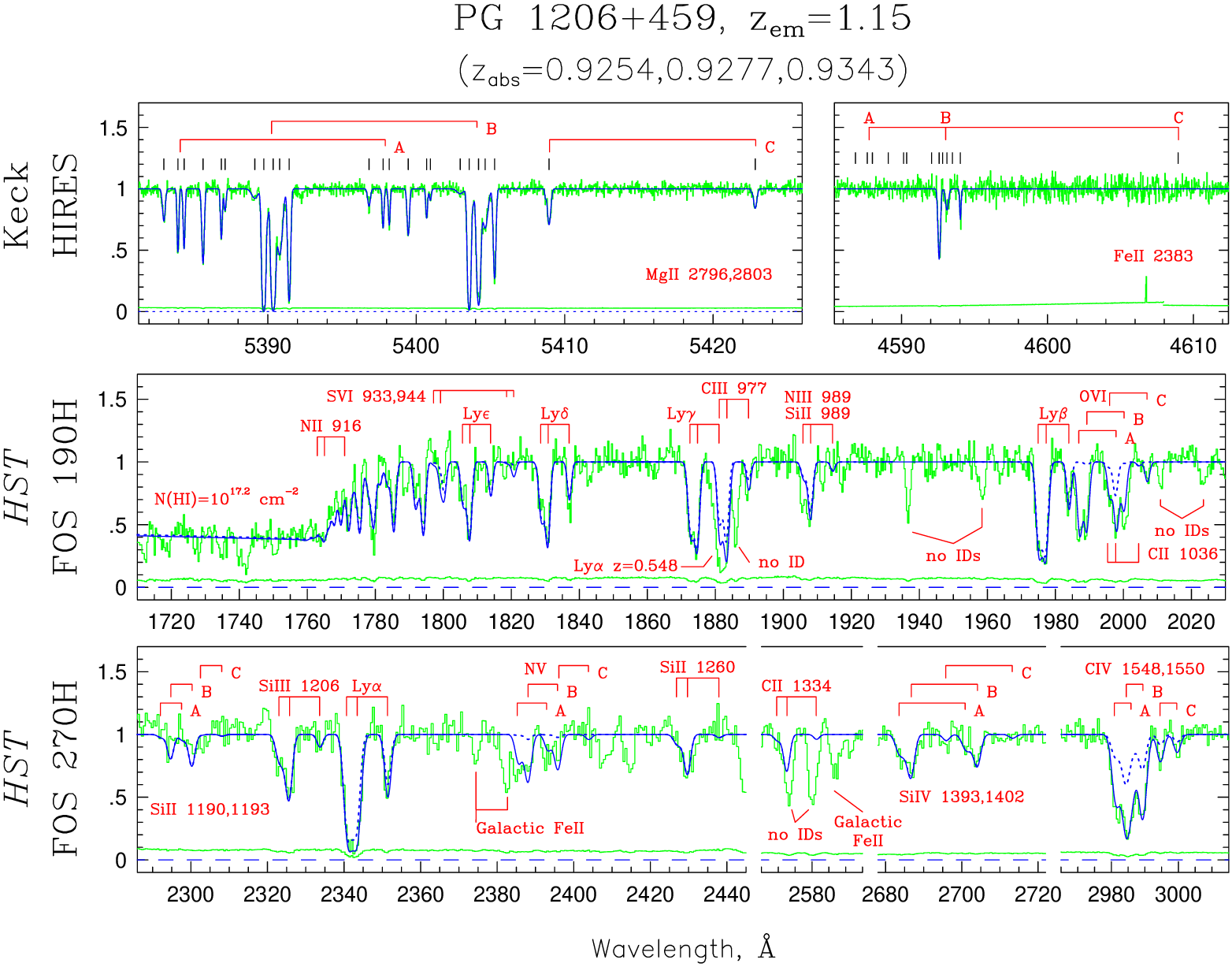}{8.0in}{90.}{80.}{80.}{271.5}{0}
\caption
{ (top) The HIRES/Keck spectra  of the {\MgIIdblt} doublet and {\FeII}
$\lambda 2600$ transition.  --- (middle and lower) The FOS/{\it HST\/}
spectrum from the Key Project database. See text for details.  }
\end{figure}

Based  upon  a  highly  significant  correlation  between  the  {\CIV}
equivalent widths and the {\MgII} kinematics, it is inferred that most
intermediate  redshift  galaxies  have  multiphase  gaseous  structures
(Churchill  et~al.\  1999, 2000).   The  low  ionization gas  is  in
multiple, narrow  components, $\left< b \right>  \simeq 5$~{\kms}, and
the high  ionization gas  is kinematically spread  out with  $\left< b
\right> \simeq  70$~{\kms} (using the doublet ratio  method).  This is
an effective velocity  dispersion, for the FOS spectra  are of too low
resolution to resolve velocity splittings below $\sim 500$~{\kms}.

\section{Case Study; The Complex Triple System at z=0.93}

The three systems at $z=0.9254$, $0.9276$, and $0.9343$ along the line
of  sight  to PG~$1206+459$  exhibit  complex  {\MgII} kinematics  and
exceptionally  strong  {\CIV},   {\NV},  and  {\OVI}  absorption.   We
investigated the ionization and  spatial distribution of these systems
using detailed photoionization models (Cloudy; Ferland 1996).

In the top  panels of Figure~2, the HIRES/Keck  spectra of the {\FeII}
$\lambda  2600$ transition and  of the  {\MgIIdblt} doublet  are shown
with a Voigt  profile model spectrum superimposed; the  ticks give the
component centers.  The systemic redshifts of the three systems, A, B,
and C, are labeled.  The lower two panels show the normalized FOS/{\it
HST\/} spectrum  (histogram) with  tuned model predictions  (not fits)
superimposed (see  Churchill \& Charton 1999).  The  dotted--line is a
single--phase model, assuming all  absorption arises due to ionization
balance in the {\MgII} clouds; a  single phase of gas fails to account
for the high ionization absorption strengths.  The solid spectrum is a
two--phase model, which allows the  higher ionization gas to reside in
a separate phase.

Based upon  the photoionization modeling, a highly  ionized phase, not
seen  in {\MgII},  is required  to  account for  the observed  {\CIV},
{\NV}, and  {\OVI} absorption.  An ``effective'' Doppler  width of $50
\leq b \leq 100$~{\kms} is consistent with the complex, blended {\CIV}
data.  The physical size of the high ionization component is less than
30~kpc, with the best values between 10 and 20~kpc.

Based upon the sizes and  effective Doppler widths, we infer that the
highly ionized  material is analogous to the  Galactic coronae (Savage
et~al.\ 1997), material stirred  up by energetic mechanical processes,
such as galactic fountains.  In this scenario, the gas is concentrated
around the  individual galaxies which  presumably provide a  source of
support, heating,  and chemical enrichment.  

It seems promising that the answer to the posed question (\S~1) may be
forthcoming when {\it  HST\/} resolves the FOS profiles  with STIS and
COS.



\acknowledgements  Thanks are due  to S.   Kirhakos, C.   Steidel, and
D. Schneider for their contributions to the work presented here.  I am
especially  grateful to  all who work to make  {\it HST\/}  a
unique platform for astronomy.

\end{document}